\documentstyle[12pt]{article}

\newcommand{\be}{\begin{equation}}
\newcommand{\ee}{\end{equation}}
\newcommand{\bea}{\begin{eqnarray}}
\newcommand{\eea}{\end{eqnarray}}

\begin{document}

\begin{center}
\begin{large}
{\bf  Dynamical Instability of Self-Tuning Solution \\}
{\bf  with\\ }
{\bf  Antisymmetric Tensor Field\\ }
\end{large}  
\end{center}
\vspace*{0.50cm}
\begin{center}
{\sl by\\}
\vspace*{1.00cm}
{\bf A.J.M. Medved\\}
\vspace*{1.00cm}
{\sl
Department of Physics and Theoretical Physics Institute\\
University of Alberta\\
Edmonton, Canada T6G-2J1\\
{[e-mail: amedved@phys.ualberta.ca]}}\\
\end{center}
\bigskip\noindent
\begin{center}
\begin{large}
{\bf
ABSTRACT
}
\end{large}
\end{center}
\vspace*{0.50cm}
\par
\noindent
We consider the dynamical stability of a static
 brane model that  incorporates a
three-index antisymmetric tensor field and  has recently been
proposed \cite{KKL1,KKL2} as a possible solution to the cosmological
constant problem. Ultimately, we are able to establish the existence of 
time-dependent, purely gravitational perturbations.  As a consequence, the
static solution of interest is ``dangerously'' located at an unstable
saddle point.  This outcome is suggestive of a hidden fine tuning
in what is an  otherwise self-tuning model.
 \\
\newpage

\section{Introduction}\medskip
\par
One of the most serious puzzles of fundamental physics is known
as the ``cosmological constant problem'' \cite{CCP}.
The essence of this problem can be summarized as follows.
Recent cosmological observations \cite{CO} place the cosmological
constant, or effective vacuum energy density, at a very small
value in comparison to the Planck scale; in fact, $\Lambda_{obs}\approx
 10^{-120}M^{4}_{PL}$. Conversely to this observational bound, one
would naively expect quantum fluctuations in the vacuum energy
to be on the order of the Planck scale; i.e., $\Lambda_{theo}\approx
 M^{4}_{PL}$. Hence, the cosmological constant problem translates
to a hierarchical problem that requires a formidable fine tuning of
120 orders of magnitude.\footnote{The situation can be somewhat improved
if we assume that supersymmetry (or any symmetry which conspires to impose
a vanishing vacuum energy) remains unbroken at energies that are just
above the present-day accelerator limits. If this were the case,
quantum fluctuations could be as small as $10^{60}M^{4}_{PL}$.
However, there would remain at least 60 orders of magnitude
to still be explained away.}
\par 
Many interesting, varied attempts have been invoked to resolve this 
cosmological constant hierarchy; albeit, with only limited degrees
of success. Some of these have been based on, for instance,
``quintessence'' \cite{Q}, the ``anthropic principle'' \cite{A} and
a probabilistic interpretation of the universe \cite{O1}.
More recently, a program that makes use of the ``brane world''
scenario has been applied in this context. 
We will focus on this approach below. 
\par
There currently exists an abundance  of different versions and interpretations
of the brane world \cite{RUB}; however, the basic picture is fundamentally
consistent. This being that ``ordinary'' matter is trapped on a 
3+1-dimensional submanifold (or ``three brane''), whereas the graviton
and (possibly) other hypothetical fields can  propagate in a 
3+1+$n$-dimensional bulk. Note that the ``extra'' bulk dimensions are
typically, but not restrictively, compact. The current popularity
of the brane world scenario can be traced to its associations
with M-theory (of which brane worlds can arise as a low-energy limit
\cite{M}), as well as its potential resolution of various hierarchical problems
(such as that between the Planck and electroweak scales \cite{ADD}).
\par
In the context of the cosmological constant problem, it is useful
to elaborate on a  specific brane  theory; namely, the
second one proposed by  Randall and Sundrum \cite{RS}, or RS2 
as it is  commonly known. In this model, a single positive-tension
brane is coupled to anti-de Sitter gravity in a  5-dimensional
bulk. Note that, for this model, there are no other bulk fields
(besides gravity) and the extra dimension is taken to be infinite.
RS2 can lead to solutions for which the 4-dimensional (effective)
cosmological constant is vanishing. However, such a solution necessitates
a fine tuning between the brane tension ($V$) and the bulk
cosmological constant ($\Lambda$) such that $V=\sqrt{-12M^{3}\Lambda}$.
(Here, $M$ is the fundamental mass scale in five dimensions.) 
\par
It has been hoped that the inclusion of new bulk fields in
the  RS2 scenario could somehow lead to a  model that permits
``self-tuning'' solutions. That is, a model for which the brane
tension could take on an extended range of values without jeopardizing
the stability of the 4-dimensional cosmological constant. 
In this way, such a model would be stable against any radiative
corrections to the brane tension and, hence, a state of  4-dimensional
Poincare invariance would be preserved.\footnote{Many models that
support  flat-brane solutions
have been shown to support curved-brane solutions as well \cite{CB}. However,
one can  assume that the flat-brane solutions are significantly
more favorable by invoking probabilistic \cite{O1} or anthropic \cite{A}
principles.}     
\par
One such candidate for  a self-tuning model has been proposed
by a pair of groups: Arkani-Hamed, Dimopoulos, Kaloper and Sundrum
\cite{ADKS}, and Kachru, Schulz and Silverstein \cite{KSS}.
The ADKS-KSS model is essentially RS2 with a vanishing bulk
cosmological constant and with coupling to a scalar (dilaton)
field.  Both studies identified apparent self-tuning solutions; and,
 for certain choices of brane-dilaton coupling, it was also found
that curved-brane solutions are conveniently forbidden.   
However, subsequent studies \cite{SLLN1,SLLN2} revealed that
the naked singularities, which  are inherent in these 
solutions,\footnote{Actually, these singularities can  be avoided;
but only along with the unphysical consequence   of  
 $M_{PL}\rightarrow\infty$, where
$M_{PL}$ is the  4-dimensional effective Planck mass.} 
lead to inconsistencies in the 4-dimensional effective
field theory. The resolution  of these inconsistencies
necessitates that an additional brane be added at each singular point.
As each  new brane leads to additional boundary conditions,
it is not surprising that 4-dimensional Poincare 
invariance 
can only be achieved via (at least one) fine tuning.
\par
As demonstrated in Ref.\cite{CJGH},
the  failure of the ADKS-KSS model to support (legitimate)
 self-tuning solutions is really just a generic feature of
a wide class of brane models with coupling to a bulk scalar.  
Furthermore, the situation does not appear to be rectified when
higher-order curvature terms are included \cite{LZ}.
\par
Even with the issue of naked singularities put aside, the
ADKS-KSS model has been critiqued on other grounds. Binetruy et al. \cite{BCG}
 have demonstrated that the field equations
(including the ``jump'' conditions \cite{IS}) support time-dependent
perturbed modes that do {\bf not} jeopardize the
Poincare invariance on the 
brane.\footnote{Perturbations of this type
(i.e., for which the brane remains flat) should
  not be confused with perturbed  modes
that induce a finite curvature on  the brane.
As previously noted, curved-brane 
solutions may be probabilistically suppressed.}  
Such perturbations imply that the static ``self-tuned'' solutions
are dynamically unstable.  From this outcome, it has been further
inferred that the brane world must evolve either from or into
a singularity, the 4-dimensional Planck mass is time dependent,
and energy fails to be conserved on the brane. This
lack of stability has also been substantiated by Diemand et al.
 \cite{DMW} via an alternative (but related) approach. Furthermore, the latter
study  found analogous  instabilities arising in a self-tuning
model that has been proposed by Kehagias and Tamvakis 
\cite{KT}.\footnote{The KT model \cite{KT}  consists of 5-dimensional gravity
and a bulk scalar field, but  no brane. Rather, for a
specific choice of  dilaton potential, there exists a parameter
limit by which a flat brane is effectively realized. Even in
this limiting case, 
the solutions are notably  free of any singularities.
However, the dynamics of the brane limit  remain unclear.} 
\par
Another promising candidate for a self-tuning theory has been recently
documented by Kim, Kyae and Lee \cite{KKL1,KKL2}. Similarly to the
 ADKS-KSS case, the KKL model is essentially the RS2 scenario along
with a new bulk field. However, rather than a scalar field, Kim et
al. have  proposed
a three-index antisymmetric tensor field.  Just such a 
field has natural origins under the compactification of 11-dimensional
supergravity \cite{O2} and   has  previously been  considered, as far back as
1980 \cite{O3,O4}, in the context of the cosmological constant
 problem. 
\par
One might  anticipate that  a three-index antisymmetric tensor 
gives rise to
 an action  term of the form $H^2$ \cite{O1}
(where  $H$ represents the
four-index field-strength tensor).   Conversely to such intuition, 
the  KKL model rather contains 
an unorthodox  $H^{-2}$ term.  It  has been shown \cite{KKL2}, however,
that
a negative power of $H^{2}$  is critical to  the self-tuning properties of
the model. For such a term to make sense, $H^2$ must develop
a vacuum expectation value on the order of the fundamental mass scale.
In this way, the KKL action could perhaps represent an effective theory
that arises out of quantum gravity.
\par
Interestingly, the KKL model allows for a static  self-tuning solution
that is endowed with  both  a
 finite 4-dimensional Planck mass
and an absence of singularities  throughout the bulk.
It remains an open question, however, whether or not the KKL model
suffers from  dynamical instabilities of the type  that plague the
ADKS-KSS and KT models (as discussed above) \cite{BCG,DMW}. The purpose of the
current paper is to rigorously address this issue.
\par
The remainder of this paper proceeds as follows. In Section 2, we 
present the KKL action and corresponding field equations, after which
the static solution is discussed. This section can be regarded as a
review of Refs.\cite{KKL1,KKL2}. In Section 3, we consider
a time-dependent analysis of the  field equations. Applying a methodology
that has been  inspired by Ref.\cite{DMW}, we are indeed able to verify
the existence of stability-threatening 
perturbed modes. The physical implications of these perturbed modes
are examined in Section 4. Finally, 
Section 5 ends with a brief summary and discussion.

\section{Field Equations and Static Solution}

We begin the formal analysis by recalling the KKL action as introduced
in Ref.\cite{KKL1}. This action describes gravity and a three-index
antisymmetric tensor field $A_{MNP}$ existing in five dimensions
and coupled to a 4-dimensional domain wall or brane (which can be
positioned at $y=0$, where $y$ is the ``extra'' bulk dimension,
 without loss of generality). More specifically,
the action of interest can be expressed as:
\be
S=\int dx^4 \int dy \sqrt{-g} \left[ {1\over 2\kappa^2}R +{2\cdot 4!\over
H^2}-\Lambda-V\delta(y)\right], 
\label{1}
\ee
where $\kappa^{-2}=2M^3$ (with $M$ being the fundamental mass scale)
and where  $H^2=H_{MNPQ}H^{MNPQ}$ is the square of the field
strength for which  $H_{MNPQ}=\partial_{\left[M\right.}A_{\left.NPQ\right]}$.  
(For a brief discussion on this choice of $H^{-2}$, see Section 1.)
Note that the bulk cosmological constant $\Lambda$ and the brane
tension $V$ are assumed to be  a negative and positive  constant, respectively.
In principle, $\Lambda$ is determined by some higher-dimensional,
fundamental theory, whereas $V$ contains all the information regarding
the Standard Model physics that lives on the brane.
\par
Varying the action with respect to the metric and antisymmetric tensor
field, we obtain the following field equations:
\be
{G_{MN}\over \kappa^2}=-g_{MN}\Lambda -g_{\mu\nu}\delta^{\mu}_{M}\delta^{\nu}
_{N} V \delta (y) + 2\cdot 4! \left({8\over H^4}H_{MPQR}H_{N}{}^{PQR}
+ g_{MN}{1\over H^2}\right),
\label{2}
\ee
\be
\partial_{M}\left(\sqrt{-g}{H^{MNPQ}\over H^4}\right)=0,
\label{3}
\ee
where $G_{MN}$ is the Einstein tensor and Greek indices represent
brane coordinates.
\par
Having interest in the dynamical behavior of solutions with
4-dimensional Poincare invariance, we invoke the following
ansatz for the metric:
\be
ds^2=e^{2A(t,y)}\eta_{\mu\nu}dx^{\mu}dx^{\nu}+b^2(t,y)dy^2.
\label{4}
\ee
On the basis of earlier studies \cite{O1,O2,O3,O4}, one can  anticipate that
the four-index field-strength tensor  is expressible 
in terms    of a  single massless scalar field and 
a  purely  geometrical,  antisymmetric  tensor field.
We thus impose the following ansatz (first suggested in Ref.\cite{KKL2})
on the field strength:\footnote{Notably,
 the four-index field-strength tensor  provides a natural
way of  separating 
a  3+1-dimensional spacetime out of the 5-dimensional bulk \cite{O2}.}  
\be
H^{\mu\nu\rho\eta}={\epsilon^{\mu\nu\rho\eta}\over\sqrt{-g}}
{\partial\over\partial y} \sigma(y,t),
\label{5}
\ee
\be
H^{4ijk}={\epsilon^{4ijk}\over\sqrt{-g}}
{\partial\over\partial t} \sigma(y,t),
\label{6}
\ee
\be
H^{04\mu\nu}=0,
\label{7}
\ee
where $\epsilon^{MNPQ}$ is the four-index (contravariant) Levi-Civita
symbol, 
 Roman indices represent spatial brane coordinates,  the index
0/4 represents the coordinate $t/y$  and  all permutations have been implied.
\par
Next, we re-express the field equations (\ref{2},\ref{3}) in terms
of the complete ansatz (\ref{4}-\ref{7}). 
Straightforward but tedious calculations yield
the following equations in the bulk: 
\be
3e^{-2A}\left({\dot A}^2b^2+{\dot A}{\dot b}b\right)-6A^{\prime 2}
-3A^{\prime\prime}+3{A^{\prime}b^{\prime}\over b}=(\kappa b)^2
\left[\Lambda+{6\over f^2-h^2} +{4 h^2\over (f^2-h^2)^2} \right],
\label{14}
\ee
\be
3e^{-2A}\left(2{\ddot A}b^2+{\dot A}^2b^2+{\dot A}{\dot b}b+
{\ddot b}b \right)-6A^{\prime 2}
-3A^{\prime\prime}+3{A^{\prime}b^{\prime}\over b}=
(\kappa b)^2
\left[\Lambda+{6\over f^2-h^2}  \right],
\label{15}
\ee
\be
3b^2 e^{-2A}\left({\ddot A}+{\dot A}^2 \right)-6A^{\prime 2}
=(\kappa b)^2
\left[\Lambda+{2\over f^2-h^2} -{4 h^2\over (f^2-h^2)^2} \right],
\label{16}
\ee
\be
-3{\dot A}+3A^{\prime}{{\dot b}\over b}=4\kappa^2{\sigma^{\prime}
{\dot \sigma}\over (f^2-h^2)^2},
\label{17}
\ee
\be
\partial_{t}\left[{\sigma^{\prime}\over (f^2-h^2)^2}\right]
-\partial_{y}\left[{{\dot\sigma}\over (f^2-h^2)^2}\right]=0,
\label{18}
\ee
where $f=b^{-1}\sigma^{\prime}$, $h=e^{-A}{\dot \sigma}$
and prime/dot denotes differentiation with respect to $y/t$. 
Note that Eqs.(\ref{14}-\ref{17}) correspond to the 00, $jj$,
44 and 04 components of the Einstein equation, while Eq.(\ref{18})
is that obtained by varying the antisymmetric tensor field. 
\par
So far, we have neglected the delta-function term in Eq.(\ref{2}).
This term leads to a discontinuity in the ``warp'' function $A(t,y)$
at the brane. Consequently, we  find that the following ``jump''
condition \cite{IS} must be satisfied:
\be
A^{\prime}(y=0^{+})=-\kappa^2{V\over 6}b(y=0^{+}).
\label{19}
\ee
To obtain this form, $Z_{2}$ (i.e., reflection) symmetry   has been assumed.
\par
Before proceeding with   the dynamical  analysis, we briefly summarize
the results of Refs.\cite{KKL1,KKL2} for  a static solution
with $b(y,t)=1$. In this case, the field equations (\ref{14}-\ref{18})
and boundary  condition (\ref{19}) reduce to the following:
\be
6A^{\prime 2}_{o} +3A^{\prime\prime}_{o}=-\kappa^2\left(\Lambda+{6\over
f^2_{o}}\right),
\label{26}
\ee
\be
6A^{\prime 2}_{o} =-\kappa^2\left(\Lambda+{2\over
f^2_{o}}\right),
\label{27}
\ee
\be
A^{\prime}_{o}(y=0^{+})=-\kappa^2{V\over 6},
\label{28}
\ee
where  $f_{o}=\sigma^{\prime}_{o}$ 
and the subscript $o$ is used to denote the static solution.
\par
Given $Z_{2}$ symmetry, the 
above equations can be uniquely  solved (up to a pair of
integration constants) to yield:
\be
A_{o}(y)=-{1\over 4}\ln\left[({a\over k})cosh
(4k\left| y\right| +c)\right],
\label{20}
\ee
\be
f_{o}=\sigma^{\prime}_{o}(y)={\kappa\over  \sqrt{3} k}cosh(4k\left| y\right|
+c),
\label{21}
\ee
where $k=\kappa\sqrt{-\Lambda/6}$, 
while $a$ and $c$ are the constants of integration.
$a$ need only be restricted to having a positive value (and can be
determined in terms of the 4-dimensional effective Planck mass),
whereas $c$ can be fixed via the jump condition (\ref{28})
as follows:
\be
tanh(c)=\kappa{V\over \sqrt{-6\Lambda}}. 
\label{22}
\ee
This relation leads to the restriction $\kappa^2 V^2 < -6\Lambda$,
but the brane tension $V$ is otherwise free to adopt any positive value.
\par
From an inspection of the static solution, the desirable features of
the KKL model are clearly evident. The integration constant $c$ can
adjust itself to moderate changes in the external parameters $V$ and
$\Lambda$ (such as quantum corrections), thereby ``protecting''
the 4-dimensional Poincare invariance. Hence, the static solution 
can be classified as one of self tuning. Moreover, the KKL solution
has no singularities while still generating   a finite
value for the 4-dimensional  (effective) Planck mass.\footnote{The 
 Planck mass
$M_{PL}$ can be directly evaluated via $M_{PL}^2=
2M^3\int^{\infty}_{0} exp\left[2A_{o}(y)\right] dy$. This has
been shown to be a  finite quantity \cite{KKL2}, 
as long as $\sqrt{k/a}$ is finite.}

\section{Linearizing the Field Equations}

To examine the time-dependent behavior of  this model, it is first
convenient to  linearize the  relevant equations 
about the static solution. Following Diemand et al. \cite{DMW},
we now express the metric functions and scalar $\sigma$   as follows:
\be
A(t,y)=A_{o}(y)+\delta A(t,y),
\label{23}
\ee
\be
b(t,y)=1+\delta b(t,y),
\label{24}
\ee
\be
\sigma(t,y)=\sigma_{o}(y)+\delta\sigma(t,y).
\label{25}
\ee
\par
To first order in ``$\delta$'', the bulk field equations (\ref{14}-\ref{18})
 take on the
following form:
\be
12A_{o}^{\prime}\delta A^{\prime}+3\delta A^{\prime\prime}
-3A_{o}^{\prime}\delta b^{\prime}=-\kappa^2\left[2\delta b \left(
\Lambda+{12\over f_{o}^2}\right)-{12\over f_{o}^3}\delta\sigma^{\prime}
\right],
\label{29}
\ee
\be
e^{-2A_{o}}\left(2{\ddot \delta A} + {\ddot \delta b} \right)=0,
\label{30}
\ee
\be
-3e^{-2A_{o}} {\ddot \delta A} +
12A_{o}^{\prime}\delta A^{\prime}=-\kappa^2\left[2\delta b \left(
\Lambda+{4\over f_{o}^2}\right)-{4\over f_{o}^3}\delta\sigma^{\prime}
\right],
\label{31}
\ee
\be
3{\dot \delta A}^{\prime} -3A_{o}^{\prime}{\dot \delta b}=-{4\kappa^2
\over f_{o}^3} \delta {\dot \sigma},
\label{32}
\ee
\be
 {\dot \delta b}-{1\over f_{o}} {\dot \delta \sigma}^{\prime}
 + {f_{o}^{\prime}\over f_{o}^{2}} {\dot \delta \sigma}=0.
\label{33}
\ee
Note that Eq.(\ref{30}) corresponds to the difference between
Eq.(\ref{15}) and Eq.(\ref{14}).  The first-order jump
condition (\ref{19}) can now be expressed as:
\be
\delta A^{\prime}(y=0^{+})= A_{o}^{\prime}(y=0^{+})\delta b(y=0^{+}),
\label{34}
\ee
where we have also made use of Eq.(\ref{28}).
\par
For the sake of simplicity, let us now assume that the time-dependent
perturbations are linear in $t$. Hence, Eq.(\ref{30}) and the
first term in Eq.(\ref{31}) can be disregarded.
\par
It is convenient to define the following combinations:
\be
\Psi\equiv \delta A^{\prime} -A_{o}^{\prime}\delta b +{4\over 3}\kappa^2
{\delta\sigma\over f_{o}^3},
\label{35}
\ee
\be
\Phi\equiv \delta b -{\delta \sigma^{\prime}\over f_{o}} +
{f_{o}^{\prime}\over f_{o}^2}\delta\sigma.
\label{36}
\ee
With  these definitions and the following useful result
(cf. Eqs.(\ref{26},\ref{27})):
\be
A_{o}^{\prime\prime}=-{4\over 3}\kappa^2 f_{o}^{-2},
\label{37}
\ee
 the first-order bulk  equations (\ref{29},\ref{31}-\ref{33})
can be rewritten
as follows:
\be
4A_{o}^{\prime}\Psi + \Psi^{\prime}- 4A_{o}^{\prime\prime}\Phi=
-4{A_{o}^{\prime}A_{o}^{\prime\prime}\over f_{o}} \delta\sigma
-{A_{o}^{\prime\prime} f_{o}^{\prime}\over f_{o}^2}\delta\sigma,
\label{38}
\ee
\be
4A_{o}^{\prime}\Psi - A_{o}^{\prime\prime}\Phi=
-4{A_{o}^{\prime}A_{o}^{\prime\prime}\over f_{o}} \delta\sigma
-{A_{o}^{\prime\prime} f_{o}^{\prime}\over f_{o}^2}\delta\sigma,
\label{39}
\ee
\be
{\dot\Psi}=0,
\label{40}
\ee
\be
{\dot \Phi}=0.
\label{41}
\ee
\par
By taking the time derivatives of  Eq.(\ref{38}) and Eq.(\ref{39}),
we are able to deduce that at least one
of  ${\dot\delta\sigma}=0$ and: 
\be
A_{o}^{\prime}=-{1\over 4}{f_{o}^{\prime}\over f_{o}}
\label{42}
\ee
must be  valid. Regardless of the former, the latter can be readily
verified by way of the static solution (\ref{20},\ref{21}).
Hence, the first-order bulk equations reduce to the simplified form:
\be
4A_{o}^{\prime}\Psi + \Psi^{\prime}= 4A_{o}^{\prime\prime}\Phi,
\label{43}
\ee
\be
4A_{o}^{\prime}\Psi = A_{o}^{\prime\prime}\Phi,
\label{44}
\ee
\be
{\dot\Psi}=
{\dot \Phi}=0.
\label{45}
\ee
\par
As an important aside, we point out that the combinations $\Psi$ and
$\Phi$ possess a special property. Namely, if we consider
``physical'' diffeomorphisms
(i.e., those  for which   the background  metric is invariant), 
then $\Psi$ and
$\Phi$ can be shown to be invariant under  such
transformations. To explicitly  demonstrate this property, we first note
 that an infinitesimal diffeomorphism of this type can be described by 
a Lie derivative with respect to some vector $X^{M}$.
The first-order perturbations are then expected to transform
as follows \cite{DMW}:
\be
\delta A \rightarrow \delta A+ A_{o}^{\prime}X^{4},
\label{46}
\ee
\be
\delta b \rightarrow \delta b+ (X^{4})^{\prime},
\label{47}
\ee
\be
\delta \sigma \rightarrow \delta \sigma + \sigma_{o}^{\prime}X^{4}.
\label{48}
\ee
By virtue of these  relations  and Eqs.(\ref{35}-\ref{37}),
it is now evident that $\Psi\rightarrow\Psi$ and
$\Phi\rightarrow\Phi$. The significance of this  invariant behavior
is as follows. If the first-order field equations (\ref{43}-\ref{45})
give rise to non-vanishing perturbations, these modes can {\bf not}
be locally transformed away by a physical diffeomorphism.\footnote{That is,
we are justified in replacing $\delta b$, $\delta A$ and $\delta\sigma$
with their gauge-invariant forms. These forms are explicitly given
in Ref.\cite{DMW} and henceforth implied.} 
\par
It is a straightforward process to solve Eqs.(\ref{43}-\ref{45}),
which yields:
\be
\Psi=Ce^{12A_{o}},
\label{49}
\ee
\be
\Phi=4{A_{o}^{\prime}\over A_{o}^{\prime\prime}} C e^{12A_{o}},
\label{50}
\ee
where $C$ is some constant.
We can now eliminate $\Psi$ and $\Phi$ from Eqs.(\ref{35},\ref{36})
and, thus, obtain a pair of differential equations with respect
to the perturbations $\delta b$, $\delta A$ and $\delta\sigma$.
\par
Since  linearity  in $t$ has  been assumed,
 the perturbations can appropriately be expressed as follows:
\be
\delta b(y,t) =k_{b}(y) + h_{b}(y)t,
\label{51}
\ee 
\be
\delta A(y,t) =k_{A}(y) + h_{A}(y)t,
\label{52}
\ee
\be
\delta \sigma(y,t) =k_{\sigma}(y) + h_{\sigma}(y)t.
\ee
The above  forms  allow us to
re-express  Eqs.(\ref{35},\ref{36}) in the following manner:
\be
k_{A}^{\prime}-A_{o}^{\prime}k_{b}-
{A_{o}^{\prime\prime}\over f_{o}}k_{\sigma}= Ce^{12A_{o}},
\label{54A}
\ee
\be
f_{o}k_{b}-k_{\sigma}^{\prime}-4A_{o}^{\prime}k_{\sigma}=4C f_{o} 
{A_{o}^{\prime}
\over A_{o}^{\prime\prime}}e^{12A_{o}},
\label{55A}
\ee
\be
h_{A}^{\prime}-A_{o}^{\prime}h_{b}-
{A_{o}^{\prime\prime}\over f_{o}}h_{\sigma}= 0,
\label{54}
\ee
\be
f_{o}h_{b}-h_{\sigma}^{\prime}-4A_{o}^{\prime}h_{\sigma}=0,
\label{55}
\ee
where we have also made use of Eqs.(\ref{37},\ref{42},\ref{49},\ref{50}).
\par
Since the interest of this paper is on the possibility of
 time-dependent perturbations,
we will restrict considerations to  the last 
pair of differential  equations (\ref{54},\ref{55}).  
What is of issue is
the existence of solutions that also  satisfy  $Z_{2}$  symmetry and
 the  jump condition at the brane.  By way of Eq.(\ref{34}),
these  boundary   conditions translate to:
\be
h_{A}^{\prime}(y=0^{+})=A_{o}^{\prime}(y=0^{+})h_{b}(y=0^{+}).
\label{56}
\ee
\par
It is clear that an appropriate solution can be obtained when  
$h_{b}=h_{\sigma}=h_{A}^{\prime}=0$.  That is:
\be
\delta b=\delta\sigma=0,
\label{59}
\ee
\be
\delta A= ht,
\label{60}
\ee
where $h$ is a constant. This solution describes  gravitational
perturbations in complete analogy with those identified
by Binetruy et al. \cite{BCG} (with regard to the ADKS-KSS model
\cite{ADKS,KSS}) and Diemand et al. \cite{DMW} (with regard
to the ADKS-KSS and KT \cite{KT} models).  As discussed in these
references, such  time-dependent perturbed modes lead
to dynamic instabilities in the otherwise static brane world.
We investigate this further in the section to follow. 

\section{Physical Interpretation}

For the sake of clarity, let us now consider the physical
implications of the dynamical perturbations as identified
in the prior section. The metric warp function  is now
revised from its static form (\ref{20}) in the 
following manner:
\be
A_{o}(y,t)=-{1\over 4}\ln\left[({a\over k})cosh
(4k\left| y\right| +c)\right] +ht,
\label{111}
\ee 
whereas the scalar function retains its static form of Eq.(\ref{21}).
Recall that $h$ is some constant parameter, $k=\kappa\sqrt{
-\Lambda/6}$, $c$ is fixed by the jump condition (\ref{22}),
and $a$ can be determined in terms of the effective Planck mass.
\par
The parameter $h$ significantly measures the
proximity of any given solution, in the ``family'' of
solutions,  to the static one.
That is, $h$ can be regarded as a continuous 
parameter for which zero  is just one particular value.
Even if one starts arbitrarily close to the static solution,
the  metric   must inevitably reach a singularity at
temporal infinity. This behavior has been interpreted  
in Ref.\cite{BCG} as the static solution being a saddle
point that  is necessarily  unstable to the smallest of perturbations.
\par
Unlike the analogous analysis  for the ADKS-KSS model,
we find no singularities in the perturbed geometry 
\cite{BCG},\footnote{That is,
in the ADKS-KSS model,
dynamical perturbations give rise to singularities in the geometry
at finite values of time.  This is not the case for the KKL model
nor for the KT model \cite{DMW}.} 
 except
for a ``big bang'' in the distant past ($t\rightarrow-\infty$) 
and a ``big crunch'' in the distant future  ($t\rightarrow\infty$).
In this sense, the situation for the KKL action can be deemed an improvement.
However, just as for the ADKS-KSS theory, an observer on
the brane will perceive a universe that does not have Poincare
invariance.  Rather, she will detect a scale factor $e^{2A(y=0,t)}$
that grows or shrinks in time.
\par
Let us now consider implications with regard to the 4-dimensional
effective Planck mass. 
We can evaluate $M_{PL}$ in terms of the fundamental mass $M$ 
by explicitly performing the $y$ integration in the action
(\ref{1}).  This yields:
\bea
M^2_{PL}&=& 2M^3\int^{\infty}_{0} e^{2A(y,t)}dy \nonumber \\
&=& e^{2ht}\left[ 2M^3\int^{\infty}_{0} e^{2A_o(y)}dy\right],
\label{222}
\eea
where the quantity in the square brackets is the static analogue
(which can be evaluated in terms of elliptic integrals
and has been shown to be finite \cite{KKL2}).
Hence,  the strength of gravity  has an undesirable
time dependence; directly in conflict with all
experimental observation.
Of course, one could suppress this effect  by fixing $h$ to be zero
or at least sufficiently small. However, this is just the type of 
fine-tuning requirement
that so-called self-tuning models profess to avoid.
That is, there is no known mechanism by which  $h=0$
can be preferentially chosen ({\it a priori}) over any other value.
\par
A final comment regarding energy conservation (or lack thereof)
is in order. This was found to be violated for the ADKS-KSS model
by virtue of the ``physical'' brane tension\footnote{In this
context, physical translates to the brane tension
as measured by an observer on the brane.} being a function
of the perturbed dilaton \cite{BCG}. (In this model, the dilaton, as well the
warp function, supports time-dependent perturbations.)
Conversely, energy should be conserved in the  perturbed KKL theory,
given that the brane tension ($V$) is decoupled from the antisymmetric 
tensor field. This would likely not be the case if we had
considered a more general (more realistic ?) brane-tension term
of the form $FV$, where $F=F(H^2)$.

\section{Conclusion}
In the preceding paper, we began the analysis by reviewing 
a self-tuning brane model that had recently been proposed by Kim,
Kyae and Lee \cite{KKL1,KKL2}. A formulation  of the static
solution revealed  how 4-dimensional Poincare invariance
can be maintained without a fine-tuning of the external parameters.
Unlike  similarly proposed self-tuning scenarios,  the  KKL static 
solution has 
no awkward singularities to be dealt with.
\par
The analysis continued with an investigation into  the dynamical behavior
 of the KKL model.  In particular, we
 linearized the  relevant  field equations  
about the static solution and then considered first-order perturbations
for which the brane remains flat.
 It was shown that the complete first-order
system (field equations and jump condition) supports a strictly
gravitational perturbation that is linear in time.  
Such a perturbed mode is known to  induce dynamical
instability in  the  brane world scenario \cite{BCG,DMW}. Other detrimental
 consequences, as we have explicitly demonstrated, 
include a time-dependent Planck mass  and a violation
of Poincare invariance as seen by an observer on the brane.   
\par
To express the problem from a different perspective, 
the static solution can be
viewed as a special member of a family of  flat-brane solutions;
these being  parametrized by $h$, where $A(y,t)=A_{o}(y)+ht$ is the warp
function. Since there is no known mechanism by which one can set 
$h=0$ {\it a priori}, it is necessary, after all, to  fine tune an
external  parameter
in the KKL model.
\par
 Once again, in the context of the cosmological constant problem,
 an apparent  self-tuning solution  has been thwarted.
In spite of this outcome, the KKL model is an intriguing approach to
the cosmological constant problem and deserves further investigation.

\section{Acknowledgements}
\par
The author  would like to thank  V.P.  Frolov  for helpful
conversations. 
  \par\vspace*{20pt}


\end{document}